\documentclass[a4paper]{jpconf}

\usepackage{graphicx}
\usepackage{amsmath}
\usepackage{amsfonts}
\usepackage{amssymb}

\begin{document}
\title{Structure formation with causal bulk viscosity}

\author{Giovanni Acquaviva$^{1}$, Anslyn John$^{2}$ and Aur\'{e}lie P\'{e}nin$^{3}$}

\address{$^{1}$ Institute of Theoretical Physics, Faculty of
Mathematics and Physics, Charles University in Prague, 18000 Prague, Czech Republic \\
$^{2}$ Department of Mathematics, Rhodes University, Grahamstown 6139, South Africa \\
$^{3}$ Astrophysics and Cosmology Research Unit, School of Mathematical Sciences, University of KwaZulu-Natal, Durban 4041, South Africa}

\ead{gioacqua@utf.troja.mff.cuni.cz, a.john@ru.ac.za, aurelie.c.penin@gmail.com}

\begin{abstract}

Dissipative features of dark matter affect its clustering properties and could lead to observable consequences for the evolution of large--scale structure. We analyse the evolution of cold dark matter density perturbations allowing for the possibility of bulk viscous pressure in a causal dissipative theory. Our analysis employs a Newtonian approximation for cosmological dynamics and the transport properties of bulk viscosity are described by the Israel--Stewart theory. We obtain a third order evolution equation for density perturbations. For some parameter values the density contrast can be suppressed compared to results obtained in the $\Lambda$CDM scenario. For other values causal bulk viscous dark matter can exhibit an enhancement of clustering. 

\end{abstract}

\section{\label{intro}Introduction} 

The universe evolved from a hot, dense, homogeneous initial state to its present cool, diffuse and inhomogeneous state during a period of $13.8$ billion years \cite{ade2016planck}. The origin of large scale structure is described by the gravitational instability paradigm \cite{weinberg2008cosmology}. In its simplest form the matter content of the universe is initially in hydrodynamic equilibrium. Density perturbations with wavelengths larger than the Jeans scale are unstable and this mechanism is thought to be responsible for the development of large scale structure and eventually galaxies. A universe consisting only of baryonic matter requires more than 14 billion years to produce the observed distribution of galaxies via the gravitational instability mechanism. Large amounts of cold dark matter (CDM) are required to salvage the structure formation scenario. Further arguments for the existence of dark matter follow from the rotation curves of galaxies \cite{persic1996universal}, lensing by clusters \cite{clowe2006direct} and the power spectrum \cite{ade2016planck} of the cosmic microwave background (CMB). The $\Lambda$CDM model includes a cosmological constant to account for the late time acceleration of the expansion \cite{peebles2003cosmological}. 

Despite its great successes the $\Lambda$CDM model encounters some problems when addressing the question of structure formation. Persistent discrepancies exist between dark matter N--body simulations and observations. These include the ``missing satellites'' problem where simulations overproduce the number of satellite galaxies \cite{Moore1999} and the ``core--cusp'' problem  where dwarf galaxies have density profiles much flatter at the center than are predicted by simulation \cite{Donato2009}.

Proposed solutions to these problems involve including baryonic feedback \cite{Brooks2013,Zolotov2012}, the warm dark matter (WDM) hypothesis \cite{viel2013warm} or extensions to general relativity i.e. modified gravity \cite{Moffat2006,Capozziello2007,BOHMER2008}. Including dissipative features -- such as viscosity -- in CDM is another approach \cite{Li2009}. This can suppress small scale structures in comparison to the $\Lambda$CDM scenario and thus help to alleviate the tension between theory and observation.

Several authors have included bulk viscosity in cosmological models \cite{Li2009, Velten2013}. Given that dark matter empirically doesn't interact with normal matter (baryons and radiation), bulk viscosity is used as a phenomenological model of possible self--interaction and consequent dissipation within the dark fluid itself \cite{Velten2013}. Most of the literature in this context invokes the Eckart theory \cite{Eckart1940}, where systems relax to equilibrium instantaneously. Here bulk viscous pressure perturbations travel infinitely fast and violate causality. Even though the Israel--Stewart theory \cite{Israel1979} represents a physically more robust way of of modelling dissipation (it is generically stable and causal), it is however more difficult to treat analytically.
 
In this paper we allow dark matter to have a bulk viscous pressure with a finite relaxation time, $\tau$, by using the causal Israel--Stewart transport equation. We determine the growth of dark matter density perturbations in this model and compare our results to the $\Lambda$CDM and Eckart models.

\section{Viscous cosmology}

To tackle the problem of structure formation with viscous dark matter we adopt a minimal model where the universe is comprised solely of spatially homogeneous and isotropic dark matter. The matter obeys Hubble's law i.e. $ \mathbf{u} = H(t) \mathbf{r} $ and cosmic dynamics is governed by the Navier--Stokes equations viz.   
\begin{eqnarray}
\frac{ \partial \rho}{ \partial t } + \nabla \cdot \left( \rho \mathbf{u} \right) &=& 0 \label{navier1} \\
\rho \left( \frac{\partial}{\partial t} + \mathbf{u} \cdot \nabla \right) \mathbf{u} &=& - \nabla p - \nabla \Pi + \rho \nabla \Phi \label{navier2} \\
\nabla^2 \Phi &=& 4 \pi G \rho. \label{navier3}
\end{eqnarray}
Dark matter is thus modelled as a self--gravitating fluid with mass density, $\rho$, velocity $\mathbf{u}$, pressure, $p$, and bulk viscous pressure, $\Pi$. The gravitational potential, $\Phi$, is determined by Poisson's equation. This Newtonian treatment is a good approximation for non--relativistic matter on sub--horizon scales and at late times. 

In the Eckart theory the bulk viscous pressure obeys the constitutive relation
\begin{equation}
\Pi = - \zeta \nabla \cdot \mathbf{u}
\end{equation}
where $\zeta$ is the coefficient of bulk viscosity \cite{belinskii1975influence} viz. 
\begin{equation}
\zeta = \zeta_{0} \left( \frac{\rho}{\rho_{0}} \right)^{s}.
\end{equation}
and quantities evaluated at the present time are indicated with the subscript $0$. The exponent $s$ encapsulates our ignorance about how the dissipation due to bulk viscosity arises from possible microscopic considerations. One only knows that, in the context of irreversible thermodynamics, the transport coefficients for a dissipative fluid depend on a certain power of the temperature, which in turn (due to Gibbs' fundamental relation) depends on a power of the energy density $\rho$.  Hence, the most generic choice for the bulk viscosity coefficient is $\zeta \propto \rho^{s}$.  In the absence of a microscopic connection, the exponent $s$ is effectively a free parameter to be fixed by observations.

In the Israel--Stewart theory the bulk viscosity obeys the more complicated transport equation  
\begin{equation}
\tau \dot{\Pi} + \Pi = - \zeta \nabla \cdot \mathbf{u} - \frac{\epsilon}{2} \Pi \tau \left[ \nabla \cdot \mathbf{u} + \frac{\dot{\tau}}{\tau} - \frac{\dot{\zeta}}{\zeta} - \frac{\dot{T}}{T} \right] \label{IStransport}
\end{equation}
where $\tau$ is the relaxation time, $T$ the temperature and the overdot represents time derivatives in comoving coordinates. In the Newtonian limit this operator reduces to the convective derivative i.e. $\frac{D}{Dt} = \frac{\partial}{\partial t} + \mathbf{u} \cdot \nabla$.

The dimensionless bookkeeping parameter, $\epsilon$, only assumes the values $0$ or $1$. It was artificially introduced into Eqn. \eqref{IStransport} as a way to later distinguish between terms arising from the full Israel--Stewart theory (where $\epsilon = 1$) and the truncated Israel--Stewart (TIS) model (where $\epsilon = 0$). The TIS model is a commonly used approximation of Eqn. \eqref{IStransport} because, while keeping the causal character of the full theory, it is analytically simpler.  However, it has been shown \cite{Maartens1996, zimdahl1996bulk} that the truncated version agrees with the full theory only in very specific cases.  Physically speaking, the condition for the validity of the truncation translates to a condition on the relative magnitude between the bulk pressure and the energy density of the fluid. In a cosmological setting, and specifically in a background (non--perturbative) analysis, the problem is that this condition might hold in some cosmic epoch but not necessarily at all times.  In general, however, it is expected that the truncated and full theories agree in near--equilibrium situations.


A crucial difference between these two theories of dissipative phenomena is that in Israel--Stewart theory, systems have a finite relaxation time, $\tau$. By contrast, in Eckart theory, systems relax instantaneously, i.e. $\tau= 0$.  This implies that bulk viscous pressure perturbations travel infinitely fast as $ c_{b}^{2} = \frac{\zeta}{ (\rho + p) \tau}.$ where $c_{b}$ is the sound speed of bulk viscous pressure perturbations. Causality is restored in Israel--Stewart theory.
 
Perturbing and linearising the system \eqref{navier1} - \eqref{navier3} leads to an evolution equation for the density contrast, $\delta \equiv \delta \rho / \rho$, which is \textit{valid for any pressure source} viz. 
\begin{equation}
\ddot{\delta} + 2 H \dot{\delta} + \left( \frac{c_{s}^{2} k^2}{a^2} - 4 \pi G \rho \right) \delta = - \frac{k^2}{a^2 \rho} \delta \Pi.
\end{equation}
We recover the standard cosmology for inviscid pressure when $\delta \Pi = 0$. For non--expanding fluids ($a=1, H=0$) we recover the original Jeans instability criterion viz. $k^2 < \frac{4 \pi G \rho}{c_{s}^{2}}$, where $c_{s}^{2} = \frac{dp}{d \rho}$ represents the adiabatic sound speed.
Including the causal, Israel--Stewart transport equation we obtain a third order evolution equation:
\begin{align}\label{ppprime}
 H \tau\, a^3\delta''' &+ \Big\{ \Big[ 3(\epsilon - q) + 1 \Big] H \tau + 1 \Big\}\, a^2\, \delta'' \nonumber\\
 &+ \left\{ \left[ \left( 3\epsilon (2-q) + j - 3 q - 4 \right) - \frac{4\pi G \rho}{H^2} + \epsilon \frac{k^2\, \Pi}{a^2H^2\rho}  \right] H \tau + (2-q) + \frac{k^2 \zeta}{a^2H\rho} \right\}\, a\, \delta'\nonumber\\
 &+\left\{ \left[ \frac{4\pi G \rho}{H^2}(4-3\epsilon) \right] H \tau + \frac{k^2}{a^2H^2 \rho} \Big( (s-1) \Pi - 3H\zeta \Big) - \frac{4\pi G \rho}{H^2} \right\}\, \delta = 0\ ,
\end{align}
where primes denote derivatives with respect to the scale factor. In this equation $q$ and $j$ are the deceleration and jerk parameters respectively viz. $q = -\ddot{a} a \dot{a}^{-2}$ and $j = - \dddot{a} a^{2} \dot{a}^{-3}$.
When $\tau \rightarrow 0$ (or equivalently $c_b^2\rightarrow\infty$) Eq.\eqref{ppprime}  reduces to Eckart's form \cite{Velten2013}.  The non-viscous $\Lambda$CDM case is recovered when $\zeta_0\rightarrow 0$. \\

\section{Analysis}

There are three characteristic timescales viz. (i) the expansion time $t_e\sim H^{-1}$, (ii) the collapse time $t_c=(4 \pi G\, \rho)^{-1/2}$ and (iii) the relaxation time $\tau=(\zeta_0\, \rho^{s-1})/(c_b^2\, \rho_0^s)$. Since we are looking at the matter-dominated era, $t_e=2/(3H)$ and, using the Friedmann equation, any ratio between $t_e$ and $t_c$ is constant. Ratios involving $\tau$ however are time dependent. In Eq.\eqref{ppprime} the relaxation time always appears in the form:
\begin{equation}
 H\, \tau = \frac{2}{3}\frac{\tau}{t_e}
\end{equation}
The deviation between IS and Eckart theories can be characterised by $\tau/t_e$. In general, when the relaxation time is non--negligible the two theories will differ. The two theories converge when $\tau/t_e\rightarrow 0$ and the condition for negligible departure from Eckart is given by
\begin{equation}\label{condition}
 \frac{\tau}{t_e}\ll 1\ \ \ \ \Rightarrow\ \ \ \ \frac{\zeta_0}{c_b^2}\sqrt{\frac{8 \pi G}{3 \rho_0}}\ a^{3\left(\frac{1}{2}-s\right)}\ll 1.
\end{equation}
Once the time-independent ratio $\zeta_0 / c_b^2$ is fixed, this condition depends strongly on $s$ which appears in the exponent of the scale factor. 
If $s<1/2$ then the condition $\tau/t_e\ll 1$ always holds for $a\ll 1$: IS and Eckart coincide at early times and the deviations can show up only at later times.
If $s>1/2$ then $\tau/t_e\gg 1$ holds at early times. Here significant deviation between IS and Eckart already appears for $a\ll 1$.
The critical value $s=1/2$ has been reported in previous studies of the background evolution of cosmological models eg. in \cite{barrow1988string,acquaviva2015nonlinear}.

\begin{figure}
\begin{center}
\includegraphics[scale=0.7]{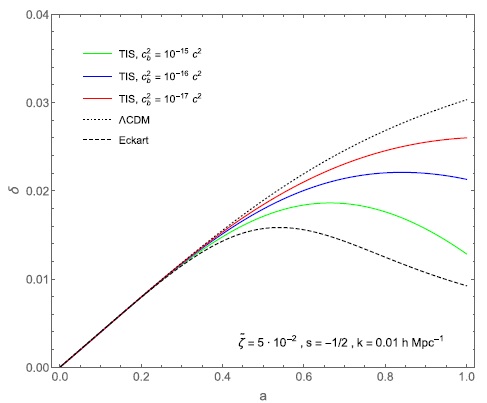}
\end{center}
\caption{\label{aaa} Evolution of the density contrast, $\delta(a)$, for $s=-1/2$, varying $c_{b}^{2}$ in TIS for $k=0.01 h$ Mpc$^{-1}$.}
\end{figure}

The density contrast evolution equation \eqref{ppprime} can be numerically integrated and a typical result is displayed in Figure \ref{aaa}. We characterise the magnitude of clustering by the amplitude of $\delta$ at late times.
The standard $\Lambda$CDM model displays the most clustering, whilst CDM with Eckart viscosity has the most strongly suppressed clustering. CDM models with truncated Israel-Stewart (TIS) viscosity lie between the two extremes. Our results indicate that Israel-Stewart viscosity models could suppress clustering at late times and alleviate the structure evolution problems. More detailed analysis can be found in \cite{acquaviva2016dark}.
For certain parameter ranges an increase in clustering compared to the non--viscous CDM case was predicted. This is surprising as a viscous fluid should be more resistant to clustering than an inviscid fluid. A possible explanation for this could be that the relevant parameter ranges i.e. values of the exponent $s$ are unphysical. The adiabatic condition viz. $\frac{D S}{D t} = 0$ needs to be modified to account for viscous energy dissipation. This is a topic for future investigation.

\section{Conclusion}

We examined the effect of bulk viscosity on the growth of dark matter perturbations in the causal Israel--Stewart dissipative theory. A third order evolution equation for the density contrast was derived and analysed both analytically and numerically. 

The numerical solutions were compared to those obtained for both the standard $\Lambda$CDM scenario as well as the non--causal Eckart theory. The deviation in behaviour of the density contrast from the $\Lambda$CDM and Eckart models depends sensitively on the exponent $s$ appearing in the bulk viscosity coefficient, $\zeta$. If $s < 1/2$ the IS models diverge from $\Lambda$CDM only at late times, whilst if $s > 1/2$ this divergence starts to occur at much earlier times. If dark matter admits bulk viscosity with coefficient $s < 1/2$ in the IS theory, the growth of structure is suppressed at late times. This may serve to alleviate the tension between the clustering of dark matter predicted by $\Lambda$CDM and that inferred by astronomical observations.

If $s > 1/2$ the IS models can exhibit greater clustering than for $\Lambda$CDM. This result is surprising as a viscous fluid should experience greater resistance to clumping than an inviscid fluid. We suspect that the parameter range might be unphysical and that this issue may be addressed by examining the energy equation for the system and establishing that the fluid's entropy is always increasing.

As expected the Eckart models approach $\Lambda$CDM in the limit of vanishing viscosity. The IS models however can still mimic $\Lambda$CDM even with non--zero viscosity if one chooses appropriate values for sound speed of bulk viscous pressure perturbations, $c_{s}$. Our analysis is based on a Newtonian approximation and solving the fully relativistic problem is the subject of future work. 
Other studies that merit attention are determining observational constraints on the viscosity parameters, as well as investigating the effects of viscosity on non--linear clustering.

\ack{AJ acknowledges financial support from Rhodes University. AP is funded by a NRF SKA Postdoctoral Fellowship. GA is funded by the grant GACR-14-37086G of the Czech Science Foundation.}

\section*{References}

\bibliography{biblio}

\end{document}